\documentclass[sigconf,screen]{acmart}

\usepackage{balance}

\AtBeginDocument{%
  }

\setcopyright{acmlicensed}
\acmDOI{10.1145/3663529.3663780}
\acmYear{2024}
\copyrightyear{2024}
\acmSubmissionID{fsecomp24ivr-p58-p}
\acmISBN{979-8-4007-0658-5/24/07}
\acmConference[FSE Companion '24]{Companion Proceedings of the 32nd ACM International Conference on the Foundations of Software Engineering}{July 15--19, 2024}{Porto de Galinhas, Brazil}
\acmBooktitle{Companion Proceedings of the 32nd ACM International Conference on the Foundations of Software Engineering (FSE Companion '24), July 15--19, 2024, Porto de Galinhas, Brazil}
\received{27-JAN-2024}
\received[accepted]{2024-04-09}

\begin{document}

\title{AutoOffAB: Toward Automated Offline A/B Testing for Data-Driven Requirement Engineering}

\author{Jie JW Wu}
\orcid{0000-0002-7895-2023}
\authornote{The author did this work before joining UBC as a postdoc.}
\affiliation{
\institution{University of British Columbia} 
\city{Kelowna}
\state{B.C.}
\country{Canada}
}
\email{jie.jw.wu@ubc.ca}

\renewcommand{\shortauthors}{Wu}

\begin{abstract}
Software companies have widely used online A/B testing to evaluate the impact of a new technology by offering it to groups of users and comparing it against the unmodified product. However, running online A/B testing needs not only efforts in design, implementation, and stakeholders' approval to be served in production but also several weeks to collect the data in iterations. To address these issues, a recently emerging topic, called \textit{offline A/B testing}, is getting increasing attention, intending to conduct the offline evaluation of new technologies by estimating historical logged data. Although this approach is promising due to lower implementation effort, faster turnaround time, and no potential user harm, for it to be effectively prioritized as requirements in practice, several limitations need to be addressed, including its discrepancy with online A/B test results, and lack of systematic updates on varying data and parameters. In response, in this vision paper, I introduce AutoOffAB, an idea to automatically run variants of offline A/B testing against recent logging and update the offline evaluation results, which are used to make decisions on requirements more reliably and systematically.

\end{abstract}



\maketitle

\section{Introduction}

Software companies are embracing a data-driven culture and are shifting from traditional requirement-based development to data-driven development and data-driven decision-making~\cite{auer2021controlled,fabijan2018online,wu2023comparison}. \textit{Online A/B testing} (also known as online controlled experiments, split tests, or randomized experiments) ~\cite{fitzgerald2017continuous,fagerholm2017right,kohavi2020trustworthy,xu2015infrastructure} is widely used to collect implicit user behavior and product effect of a given change for online and web-facing products, such as social media \cite{xu2015infrastructure}, search engines~\cite{tang2010overlapping}, social networks~\cite{xu2015infrastructure,feitelson2013development}, and web services~\cite{turnbull2019learning,pajkovic2022algorithms}. The procedure offers different product variants to different groups of users, then collects data related to the user behavior, and compares the different product variants to the unmodified product~\cite{fitzgerald2017continuous,fagerholm2017right}. The A/B test allows the gathering of information for a small but significant percentage of users for stakeholders to make decisions on whether to launch a particular variant to 100\% of users \cite{kohavi2020trustworthy,xu2015infrastructure}.

However, online A/B test suffers from several limitations. First, it takes significant development efforts to design and implement the change in the code base, with production-level standards. Second, the change will have a real impact on a relatively large group of users in the A/B test to get statistically significant A/B results. So it could affect users negatively if the change in the A/B test includes any bug or safety issue. Thus, domain owners of the products need to sign off for them to be served to a subset of users. Lastly, it typically needs several weeks to run the A/B tests to collect the data with potentially multiple iterations~\cite{kohavi2020trustworthy}. These limitations dramatically increase the time for the product team to try new ideas. 

To address these pain points, a lot of researchers have studied the emerging topic of \textit{offline A/B testing} (or offline policy evaluation, counterfactual evaluation)~\cite{joachims2016counterfactual,gilotte2018offline,gruson2019offline,reklaite2022offline,saraswat2021hybrid}. The objective of offline A/B testing is to conduct offline evaluation of a new technology by estimating from historical logged data~\cite{joachims2016counterfactual}.  A number of estimators have been developed such as importance sampling (IS)~\cite{li2011unbiased}, capped importance sampling (CIS)~\cite{bottou2013counterfactual}, normalized and capped importance sampling (NCIS)~\cite{swaminathan2015self} to reach a good balance between bias and variance~\cite{chen2023opportunities}, therefore increasing the correlation between the estimated results in offline A/B tests and the actual results in online A/B tests. 

\begin{figure*}[h]
  \centering
  \includegraphics[width=\linewidth]{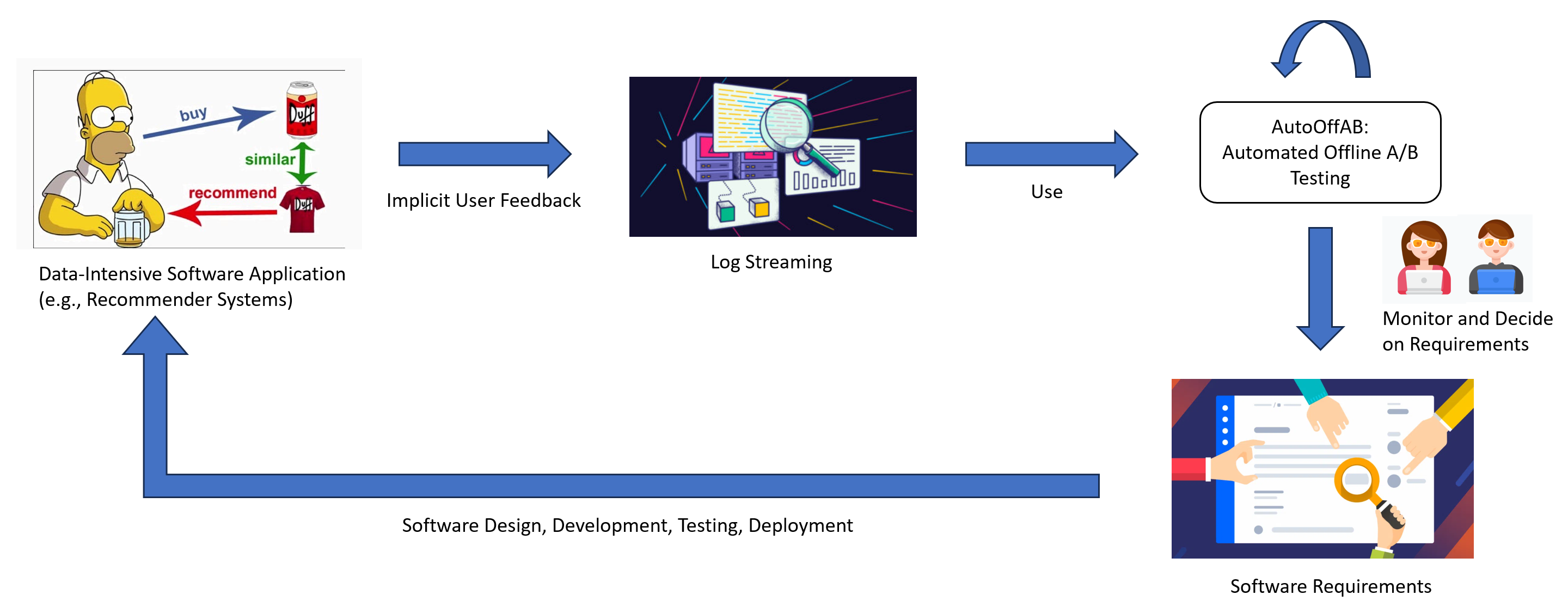}
  \caption{Visual illustration of the proposed AutoOffAB in the context of Data-Driven Requirements Engineering (DDRE) cycle~\cite{maalej2015toward}. Without this work, the offline A/B testing needs to be conducted manually by software engineers or ML scientists, which depends heavily on their individual skills. With this work, the offline A/B testing is triggered periodically. Thus, engineers or scientists could focus on monitoring and reviewing the results to be used for decisions on requirements. }
  \label{fig:ddre}
\end{figure*}

Although offline A/B testing is a promising approach due to much smaller development effort and faster turnaround time, there are still limitations for it to be reliably and effectively used in requirements engineering in practice. The offline A/B testing is a manual process that runs the offline evaluation for manually selected algorithms (or policy~\cite{gilotte2018offline}) against the one-off historical data. Therefore, there is a lack of systematic updates of offline evaluation on either 1) the updated and chosen historical data or 2) other algorithms that could be more optimal than the manually selected ones. This can lead to unreliable offline results and potentially enlarges the discrepancy between offline and online A/B test results. 

To address this limitation, in this paper, I introduce AutoOffAB, an idea to automatically generate and periodically update the offline A/B testing evaluation towards more reliable and systematic offline A/B test results for making decisions in requirements engineering. The automation produces offline evaluation as periodic updates from recent historical data rather than one-off historical data. This can prevent the \textit{outdated} evaluation results due to any recent product change. Meanwhile, the periodic automated process also generates results for modified technologies using either randomized genetic algorithms (GA)~\cite{holland1992genetic} or potentially more sophisticated methods in the future, rather than a limited set of manually selected technologies in a manual process. I believe that the results from AutoOffAB is more reliable and systematic than the current manual process so that the results and numbers can be more trusted when being prioritized in \textit{Data-Driven Requirements Engineering (DDRE)}~\cite{maalej2015toward}, as shown in Figure~\ref{fig:ddre}. More reliable numbers can also help reduce the gap between the offline A/B testing results and the online A/B testing results, which is a critical criterion for the effectiveness and usefulness of offline A/B testing. In the remaining parts of this paper, I describe the idea of AutoOffAB in detail and discuss possible ways to realize it.

\begin{figure*}[h]
  \centering
  \includegraphics[width=\textwidth]{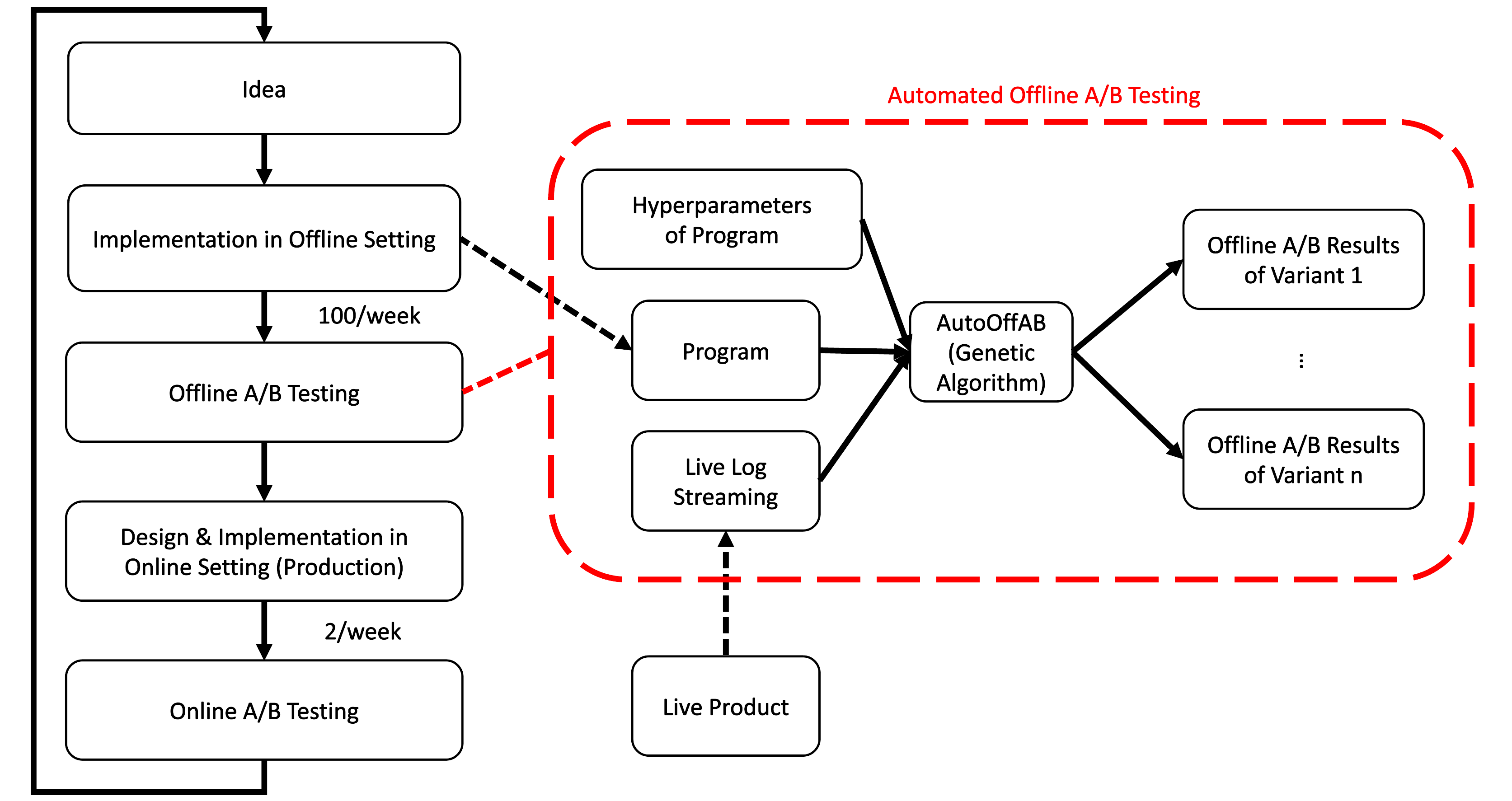}
  \caption{Visual illustration of AutoOffAB. Overall, AutoOffAB uses the program and log streaming to \textit{periodically} generate a population of variants with \textit{modified settings}, and then evaluates the variants against the \textit{updated chosen logs}.  }
  \label{fig:autooffab} 
\end{figure*}
\section{Offline A/B Testing Analysis}
\subsection{The Current State of Offline A/B Testing}
In the current software industry, offline A/B testing is a manual process conducted by software engineers or ML scientists. It has been used in different data-intensive products such as search~\cite{joachims2016counterfactual}, recommendation~\cite{saraswat2021hybrid}, ad placement~\cite{bottou2013counterfactual,joachims2016counterfactual}, etc. The steps of the offline A/B testing are described as follows. First, the software engineers or ML scientists decide what type of log data to use in the offline evaluation. Second, they select one or a few algorithms to be evaluated. The settings of an algorithm include hyperparameter values, modeling decisions, feature sets, etc. Third, they define the metrics for offline evaluation. Finally, they conduct the evaluation to generate evaluation results for each algorithm against the logged data. Each experimental result corresponds to each algorithm with its setting. Although the offline evaluation significantly reduces the turnaround time of iterating new ideas, it is assumed that the software engineers or ML scientists have full experience with the following questions:
\begin{itemize}
\item How many algorithms to evaluate?
\item How to determine the settings of these algorithms?
\item How to select the historical logs in offline evaluation?
\item How to define the cadency of running the evaluation?
\end{itemize}

However, it appears in some studies that this currently manual process has the following drawbacks:

\begin{itemize}
\item It depends solely on the engineers or scientists to decide which setting or 
parameter values of an algorithm to be used in the offline evaluation. Thus, the choice of variants and their parameter
values rely heavily on the skills of engineers, who usually have little assistance or guidance in choosing variants.
\item It is often not humanly possible to try all combinations of settings and parameter values to obtain the parameters that lead to the precise optimal result.
\item The offline evaluation is a one-off job on certain data from historical logs, but the evaluation results may be inconsistent with the data from different logs (such as most recent data, or shuffled data using certain strategies).
\end{itemize}

As an example to illustrate these issues, the offline A/B testing conducted for playlist recommendation~\cite{gruson2019offline} used 12 algorithms, with different settings on hyperparameter values, modeling decisions, training data definition, etc. Although the researchers tested 12 different settings extensively, it is still not realistic to try all possible combinations of settings to get the most optimal results. Also, the offline evaluation is a one-off job on the predefined historical logs from systems running in production. However, the product is live and is collecting data every day, with new changes in the product or users as time goes by. So the one-off results of the old historical logs may not hold the same for the recent logs. These issues are not fully explored and addressed.

\subsection{Motivation of Automated Offline A/B Testing}

My idea is to periodically trigger the offline A/B testing evaluation on the \textit{updated} logs with \textit{modified variants}, instead of manually running it as a one-off job. Intuitively, this can lead to more reliable and more comprehensive offline A/B test results to be prioritized in requirements analysis and specifications. As \textit{data} has become an intrinsic and evolving component in software development, the motivation behind this idea is to replace the manual process with the automated process for offline evaluation based on data of data-intensive applications. This idea addresses the pain points of the current manual process as mentioned above. First, it alleviates the burden of engineers or scientists to choose the settings of the algorithms that may lead to the "best" results based on their educated guesses and skills~\cite{simon2023exploring}. In fact, these educated guesses are often inaccurate, as there is evidence pointing out that these intuitions are often wrong and contradict the data from the A/B testing~\cite{kohavi2020trustworthy}. Second, even if the educated guesses are in the right direction due to the strong skillsets, the machine does a much better job than humans on trying all combinations of settings to find the precise optimal results~\cite{harman2012search,wu2023multi}. Third, reliability can be better ensured by repeating the offline evaluation on logs that are continuously updated and variants with modified settings~\cite{jia2010analysis}. 

\section{Proposed Architecture for Automated Offline A/B Testing}
\subsection{Overview}
In this paper, I present AutoOffAB to automate the manual procedure of offline A/B testing. Figure~\ref{fig:autooffab} provides a visual illustration of AutoOffAB. On the left side, the experiment lifecycle is displayed. As aforementioned, due to the considerable amount of time for collecting data on a large user base and efforts on the design and implementation, only a small number of ideas can be tested in online A/B tests in actual working environments of software companies. Therefore, offline A/B testing can be an area of high Return on Investment (ROI) if the offline A/B results have a strong alignment with online A/B results. Currently, engineers or scientists typically run a one-off offline evaluation on certain historically logged data. 

The red dashed box in Figure~\ref{fig:autooffab} illustrates how AutoOffAB works. AutoOffAB is based on three components:
\begin{itemize}
\item Program,
\item Hyperparameters of Program,
\item Log Streaming.
\end{itemize}
The program refers to the implementation of offline A/B testing. Hyperparameters of the program refer to the settings of the program that result in modified technologies such as the model hyperparameters, external and internal parameters, modeling choice, feature set, choice of algorithmic variants, etc. Log streaming is the logs collected by the live product. AutoOffAB uses the program and log streaming to \textit{periodically} generate a population of variants with \textit{modified settings}, and then evaluates the variants against the \textit{updated chosen logs}. 

Next, I discuss the following steps of AutoOffAB: 1) Hyperparameter Specification, 2) Algorithm Design of Variants Selection, 3) Evaluation of Variants. 4) Operations and Monitoring of Variants. 5) Decisions on Requirements.

\subsection{Hyperparameter Specification}
I refer to Hyperparameter Specification as specifying the hyperparameters of the program for the offline A/B evaluation. This includes the specification of all of the hyperparameters that potentially impact the evaluation results, and their valid values. These hyperparameters typically include external and internal parameter values, modeling choices, feature set, training data sources, etc~\cite{gruson2019offline}. Formally, the program $m$ of offline A/B testing contains a set of hyperparameters $P=\{p_1,...,p_n\}$, where each hyperparameter $p_i$ has a corresponding range of valid values $R_i$. So the goal in this step is to specify $P=\{p_1,...,p_n\}$ and $R=\{R_1,...,R_n\}$. Different types of hyperparameters for the program $m$ can be included, such as the one that specifies which data to use from the log streaming. 

\subsection{Algorithm Design of Variants Selection and Evaluation}
After the hyperparameters $P$ and their valid value ranges $R$ are specified, the next step is to generate and select variants $V=\{v_1,...,v_m\}$ for evaluation. Each variant $v_j$ is an implementation of the program $m$ with its assigned values for hyperparameter $P=\{p_1,...,p_n\}$. In the manual process, engineers or scientists need to use their experience and skillset to assign the hyperparameter values. In the proposed automated architecture, a genetic algorithm is used to generate a population of variants in each offline evaluation $e_k$. These variants are then selected and evaluated against the historical data $D_{e_k}$, based on the pre-defined measurement $c$ for offline evaluation. The measurement is used as a fitness function for the genetic algorithm. So, the objective of the genetic algorithm is to find a variant $v$ that maximizes $c(v,D_{e_k})$, the evaluation measurement of the given variant $v$ and data $D_{e_k}$ from the evaluation $e_k$. The genetic algorithm is chosen because of its simplicity and wide usage, but any search and optimization technique in Search Based Software Engineering (SBSE)~\cite{harman2012search} or more sophisticated methods could be used in future work. Note that if the measurement function $c$ has multiple objectives rather than one objective, multi-objective optimization such as Multi-Objective Evolutionary Algorithm (MOEA)~\cite{deb2011multi} may be used.

\subsection{Results Monitoring and Decisions on Requirements}
The last step of AutoOffAB is related to how to analyze and monitor the periodic, automated offline evaluation results, and how to use the results to make decisions and prioritizations on requirements. First, the continuous and automated evaluation results need to be monitored and analyzed to find out if there are any non-trivial findings that are worth discussing and will potentially impact requirements engineering, such as any change in evaluation results on most recent data, any abnormal results on certain models, etc. The monitoring and analysis are expected to be lightweight, without heavy efforts in data analytics. The stakeholders need to review the evaluation results and make decisions on requirements for which work items need to be prioritized based on the results. With the automated offline A/B testing results, a continuous cycle is formed, known as the Data-Driven Requirement Engineering cycle. As shown in Figure~\ref{fig:ddre}, the cycle starts with implicit feedback from users via logging. Then, the automated offline A/B testing is run to generate evaluation results. The results are monitored and analyzed by engineers or scientists to make decisions on requirements. Finally, the derived requirements are executed, developed, and launched in the development process. After the new launches, The cycle starts again with implicit user feedback from updated chosen data.

\section{Future Plans}
\textbf{Public Benchmark and Baseline Methods.} As for the future plan, firstly, a public benchmark together with baseline methods and their evaluations need to be created for the task of automating offline A/B testing. The benchmark will enable the comparison between different algorithms and will facilitate the development and evaluation of offline A/B testing. Secondly, I have formulated this problem as an optimization problem, so different optimization methods or search-based methods can be used as baselines for this problem, such as the GA-based methods mentioned in this work.

\textbf{LLM-based Automation for Offline A/B Testing.} Finally, given the recent advances in Natural Language Processing (NLP) and Large Language Models (LLM) on code generation and software engineering tasks, LLM-based approaches should be explored to create more intelligent non-trivial variants for offline A/B testing so that the capability of automated offline evaluation will go beyond changing hyperparameter values with search-based methods only. 

\section{Conclusion}
The great potential of offline A/B testing has attracted much interest from both academia and the software industry recently. The beauty of offline A/B testing lies in its support of much faster iteration of trying ideas in practice for many data-intensive ML-enabled applications such as search, recommender systems, and advertising. In this paper, I present AutoOffAB, an idea toward automated offline A/B testing. AutoOffAB automatically runs and periodically updates the offline A/B testing results, which are used to make decisions on requirements for further development. Given the importance of offline A/B testing, I argue that there should be a better presence for Software Engineering research to enable more reliable and systematic offline A/B test results via solutions like AutoOffAB. 

\clearpage
\bibliographystyle{ACM-Reference-Format}
\balance
\bibliography{sample-base}

\end{document}